\documentclass[amsmath,amssymb, aps, prb, superscriptaddress,twocolumn]{revtex4-1}

\usepackage[pdftex,colorlinks=true]{hyperref}
\hypersetup{
citecolor = blue
}

\usepackage{graphicx}
\usepackage{dcolumn}
\usepackage{bm}
\usepackage{amsmath}
\usepackage{float}
\usepackage{hyperref}
\usepackage{wrapfig}
\usepackage{amssymb}
\usepackage{enumitem}
\usepackage{bm}
\usepackage{subfigure}
\usepackage{placeins}

\usepackage[normalem]{ulem}
\usepackage{amsthm}
\usepackage{amsfonts}
\usepackage{bbm}
\usepackage{array}

\begin{document}

\title{Machine learning out-of-equilibrium phases of matter}

\author{Jordan Venderley}
\affiliation{\mbox{Department of Physics, Cornell University, Ithaca, New York 14853, USA}}

\author{Vedika Khemani}
\affiliation{\mbox{Department of Physics, Harvard University, Cambridge, MA 02138, USA}}

\author{Eun-Ah Kim}
\affiliation{\mbox{Department of Physics, Cornell University, Ithaca, New York 14853, USA}}

\date{\today}
\begin{abstract}

{Neural network based machine learning is emerging as a powerful tool for obtaining phase diagrams when traditional regression schemes using local equilibrium order parameters are not available, as in many-body localized or topological phases.  Nevertheless, instances of machine learning offering new insights have been rare up to now. Here we show that a \emph{single} feed-forward neural network can decode the defining structures of two distinct MBL phases and a thermalizing phase, using entanglement spectra obtained from individual eigenstates.
For this, we
introduce a simplicial geometry based method for extracting multi-partite phase boundaries. We find that this method outperforms conventional metrics (like the entanglement entropy) for identifying MBL phase transitions, revealing a sharper phase boundary and shedding new insight into the topology of the phase diagram. Furthermore, the phase diagram we acquire from a {\emph single} disorder configuration confirms that the machine-learning based approach we establish here can enable speedy exploration of large phase spaces that can assist with the discovery of new MBL phases. To our knowledge this work represents the first example of a machine learning approach revealing new information beyond conventional knowledge.}

\end{abstract}
\maketitle

The application of machine learning (ML)\cite{MLTrends} to central questions in the theory of quantum matter
is a rapidly developing research frontier.
So far, efforts have been two-fold, focusing on: (1) representing states compactly\cite{Chen:2017,Deng2017,Sarma2016,Fu2017} and (2) identifying and classifying different phases of matter\cite{Trebst2017,Trebst2016,Zhang2017,Wang2016,Carleo2017,Melko2017,Huber2017,ZhangZ2}.
The driving insight here is that the problems of theoretical interest are essentially those of regression in which an exponentially large amount of data must be condensed into a more accessible or meaningful form, e.g. the labeling of wavefunctions with phases. As neural networks are universal function approximators and facilitate nonlinear regression, neural network based ML can effectively distill relevant information from complex data while taking it at face value. This is particularly appealing for phases outside the traditional regression scheme where a local order parameter may not be readily available. Such phases include topological phases and out-of-equilibrium \emph{eigenstate phases} \cite{Huse13, PekkerHilbertGlass} in the context of many-body localization (MBL) \cite{Anderson58, Basko06, PalHuse, Znidaric, OganesyanHuse,  Nandkishore14, AltmanVosk}.  Alhough there has been recent progress in using ML for topological phases
\cite{Zhang2017,ZhangZ2,Ohtsuki} and MBL phases\cite{Neupert2017,Huber2017}, extracting phase boundaries in these settings has been a challenging frontier.\cite{Neupert2017,Huber2017} Moreover, the question of whether the same data and architecture can be used to discern multiple phases, especially multiple MBL phases has been unclear.  

\begin{figure}[ht]
\centering
\subfigure{
    \hspace{-10pt}\includegraphics[width=0.55\linewidth]{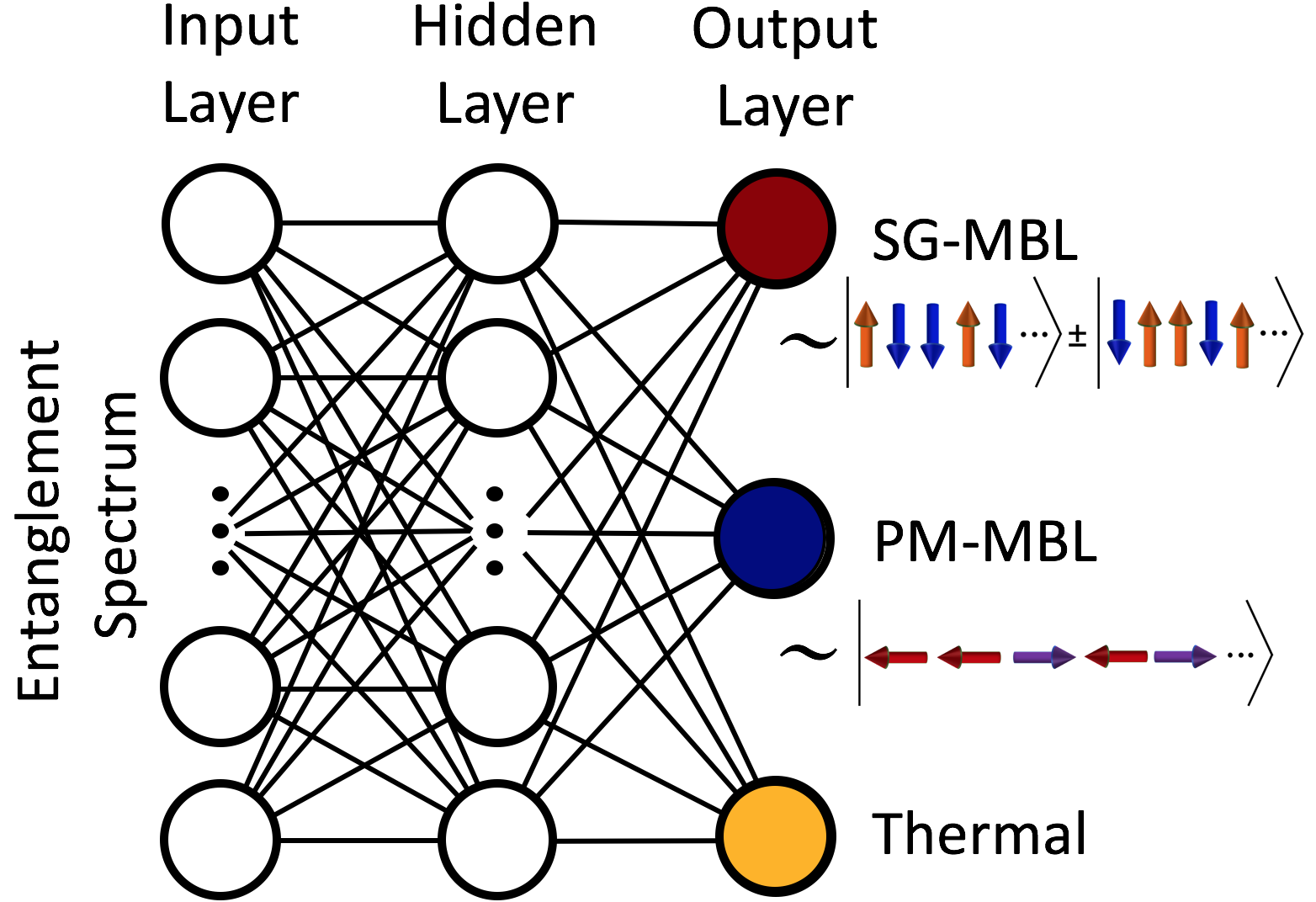}}
\subfigure{
    {\includegraphics[width=0.45\linewidth]{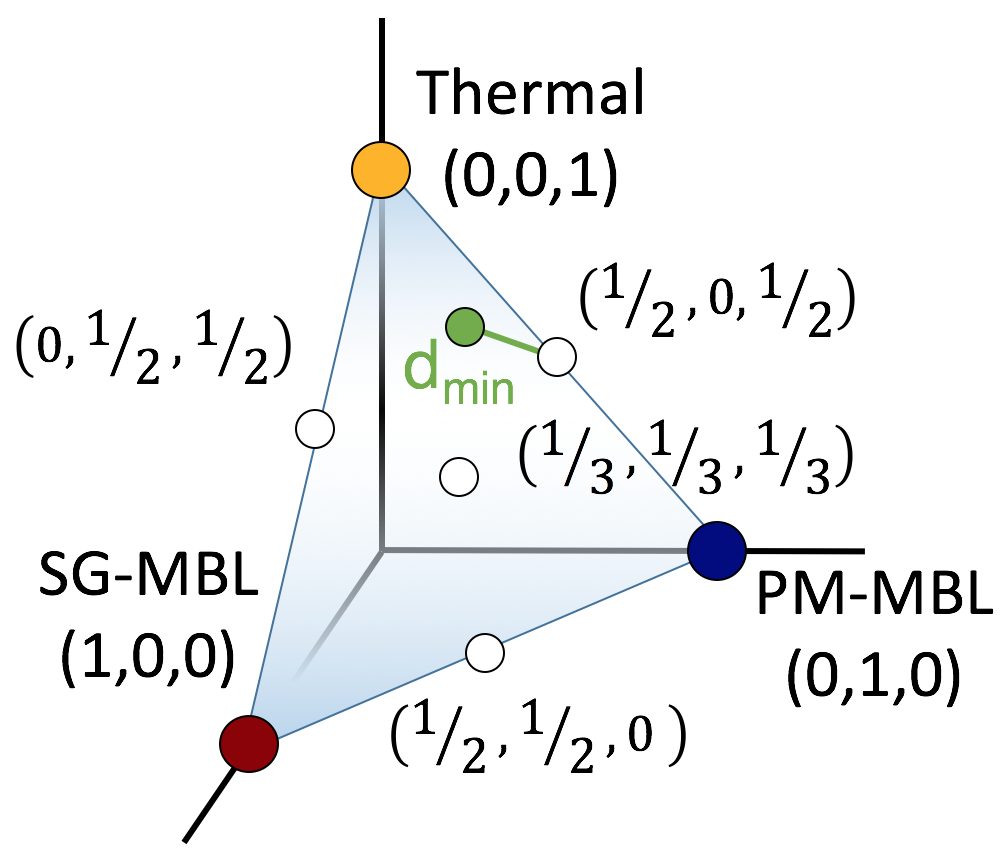}}}
\begin{picture}(0,0)
\put(-130,110){(a)}
\put(0,110){(b)}
\end{picture}
\caption{(a) A depiction of our neural network. (b) The 2-simplex codomain of our neural network outputs. 
Colored circles represent regions where the network picks a given phase with  100\% confidence, while the white circles represent regions of maximal confusion. The green point represents an example output from the neural network  with its associated $d_{\rm min}$ marked with a green line. 
}
\label{fig:NN_cartoon}
\end{figure}
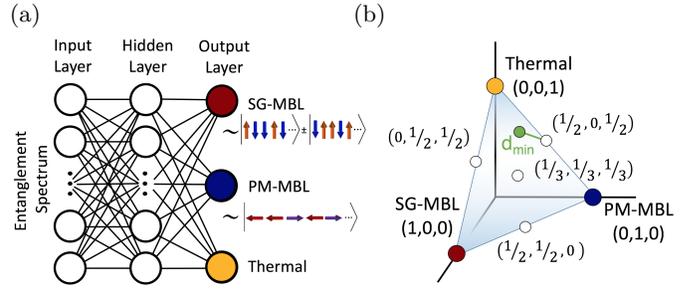

MBL generalizes the phenomenon of Anderson localization to the interacting setting, bringing out the interplay of disorder and interactions. Since MBL systems stay out of thermal equilibrium, they can display a host of rich dynamical phenomena\cite{Huse14, Serbyn13cons, Imbrie2016,VasseurRevivals,KhemaniNonLocal,Gopalakrishnan15}. Furthermore it is now known that different varieties of MBL phases (e.g. MBL paramagnets, symmetry breaking MBL phases, topologically ordered MBL phases etc.) --- each showing different patterns of order in individual highly-excited many-body eigenstates --- can be realized in a given system\cite{Huse13, PekkerHilbertGlass, Bauer13, Chandran14, Bahri15}. Some of these phases might be forbidden in equilibrium, and transitions between these distinct MBL phases 
are novel \emph{dynamical} eigenstate phase transitions that are invisible to standard thermodynamic ensembles.
With experimental realizations of novel out-of-equilibrium states in MBL settings such as time crystals\cite{Khemani15, Q, CVS, MishaTCExp, MonroeTCExp}, it is all the more important to understand the nature of these out-of-equilibrium phases and the transitions between and out of them. Moreover, we need efficient ways to study and discover new MBL phases \emph{without} an \emph{a priori} knowledge of the defining order-parameters.   

Despite extensive research~\cite{PalHuse, Luitz15, VHA,PVP,GroverCP,SerbynCriterion, ClarkBimodal,KhemaniCP, KhemaniCPQP}, a complete theoretical understanding of the transitions between different MBL phases and between the MBL and thermal phases is lacking, partially due to the absence of a comprehensive scheme for regression. Although entanglement entropy serves as a useful diagnostic of thermalization (highly excited many-body eigenstates in the thermal phase are “volume-law” entangled, while they are only area-law entangled in the MBL phase\cite{PalHuse,Bauer13}), it appears to be too aggressive a regression since it traces out important entanglement correlations. The structure of these entanglement correlations is expected to be relevant for understanding the nature of the many-body entanglement ``resonances'' that drive the transition out of the MBL phase\cite{VHA,PVP,KhemaniCPQP} and for more broadly revealing the nature of transition. Moreover the entanglement spectra of individual many-body eigenstates must encode the structure of different MBL phases, even when the defining correlation functions are not known \emph{a priori}. While there have been efforts to utilize the full entanglement spectra (ES)\cite{GeraedtsES1, GeraedtsES2}, a complete understanding for how to interpret the ES has not yet been established. 
Alternatively there have been efforts to employ neural networks to extract relevant information from entanglement spectra 
\cite{Neupert2017,Huber2017}, but it has been unclear whether ML has been able to offer any new insights thus far. 

In this work, we take a first step towards a ML assisted study of MBL phase transitions  by using a neural network to obtain a tripartite phase diagram containing two distinct MBL phases and a thermal phase. For concreteness, we work with a disordered and interacting transverse-field Ising model (TFIM) which has two distinct many-body localized phases: 
 (1) many-body localized spin-glass (MBL-SG) and (2)  many-body  localized  paramagnetic  (MBL-PM), in addition to the thermal phase. 
 Using the entanglement spectra of individual eigenstates as our only input, we are able to locate these phase boundaries with far greater precision than standard methods for studying MBL transitions.  To do this, we introduce a new geometric approach for interpreting neural network outputs for multipartite classification, and we expect this method to be of interest in its own right.  

{\it Model -- }The TFIM in the presence of disorder and interactions is a ``canonical model'' for studying novel eigenstate phases\cite{Huse13, PekkerHilbertGlass, Kjall14}. It has a well-studied non-interacting limit\cite{Fisher95}, and well-understood descriptions for the paramagnetic and spin-glass phases in the different limits of this model. An Ising self-dual variant of this model for an $L$ site chain is given by\cite{Moudgalya2017}: 
\begin{equation}
H = - \sum\limits_{i=1}^L \Big[ J_i \sigma_i^z \sigma_{i+1}^z + h_i \sigma_i^x 
 + \lambda \big(\bar{h}\sigma_i^x \sigma_{i+1}^x + \bar{J}\sigma_i^z \sigma_{i+2}^z\big) \Big],
\label{eq:model}
\end{equation}
where $\sigma_i^z$ are Pauli spin 1/2 matrices on site $i$. 
The couplings, $\{J_i\}$, and onsite fields, $\{h_i\}$, are drawn from log-normal distributions such that the standard deviation of their logarithms is $\delta(\log J) = \delta(\log h) = 1$. Note that this model is equivalent to a disordered interacting fermion model upon a Jordan Wigner transformation, where the interaction strengths are proportional to $\lambda$. Finally, $\bar{h}$ and $\bar{J}$ denote the means of $\{J_i\}$ and $\{h_i\}$\footnote{Scaling the interaction terms with $\bar{J}, \bar{h}$ ensures that 
the interactions are not overwhelmed by the non-interacting terms which can be large on account of the log-normal distributions.}. 
The means $\overline{\log J}$,  $\overline{\log h}$, 
and $\lambda$ serve as tuning parameters that can be used to drive different phase transitions. 

Let us first consider the axis $\lambda = 0$, which is equivalent to a disordered free-fermion model subject to Anderson localization and area law entanglement. From the eigenstate order perspective, there are two distinct phases with respect to the global $\mathbb{Z}_2$ Ising symmetry of the model under spin flips $P = \prod_i\sigma_i^x$: the symmetry-broken spin-glass (SG) phase for $\bar{J} >\bar{h}$ and the  paramagnetic (PM) phase for $\bar{J} <\bar{h}$. Deep in the SG phase, individual many-body eigenstates are macroscopic superpositions (i.e. Schrodinger ``cat'' states) in the $\sigma^z$ basis with localized domain walls: $|\alpha \rangle \sim |\uparrow \downarrow \downarrow \uparrow \downarrow ... \rangle \pm |\downarrow \uparrow \uparrow \downarrow \uparrow ... \rangle $, and the connected correlation function of $\sigma^z$ evaluated in each such eigenstate shows long-range order with 
\begin{equation}
\langle \alpha| \sigma_i^z \sigma_j^z|\alpha\rangle_c\equiv\langle \alpha | \sigma_i^z \sigma_j^z | \alpha \rangle - \langle \alpha | \sigma_i^z| \alpha \rangle\langle \alpha | \sigma_j^z | \alpha \rangle = \pm |c_\alpha|,  
\label{eq:SG}
\end{equation} 
$|c_\alpha|>0$ even as $|i-j| \rightarrow \infty$.
By contrast, for the equilibrium problem in the absence of disorder, a finite density of delocalized domain walls destroys long-range order at any finite temperature in 1D in accordance with  Peierls-Mermin-Wagner theorems. Thus, the SG phase in 1D furnishes an example of a model where localization enables a new form of dynamical order that is disallowed in equilibrium and invisible to the thermodynamics\cite{Huse13, PekkerHilbertGlass}. 
On the other hand, the eigenstates deep in the PM phase resemble product states in the $\sigma^x$ basis,  $| \alpha \rangle \sim | \rightarrow \leftarrow \leftarrow \rightarrow \leftarrow ... \rangle$ without LRO, i.e., 
$\langle \alpha | \sigma_i^z \sigma_j^z|\alpha \rangle_c = 0$.
The critical point between these two phases is at the Ising self-dual point, $\overline{\log J} = \overline{\log h}$, and the critical properties 
for $\lambda=0$ are described by an infinite randomness fixed point\cite{Fisher95}.  

\begin{figure}[ht]
\centering
\vspace{10pt}

\subfigure[]{
	\includegraphics[width=0.95\linewidth]{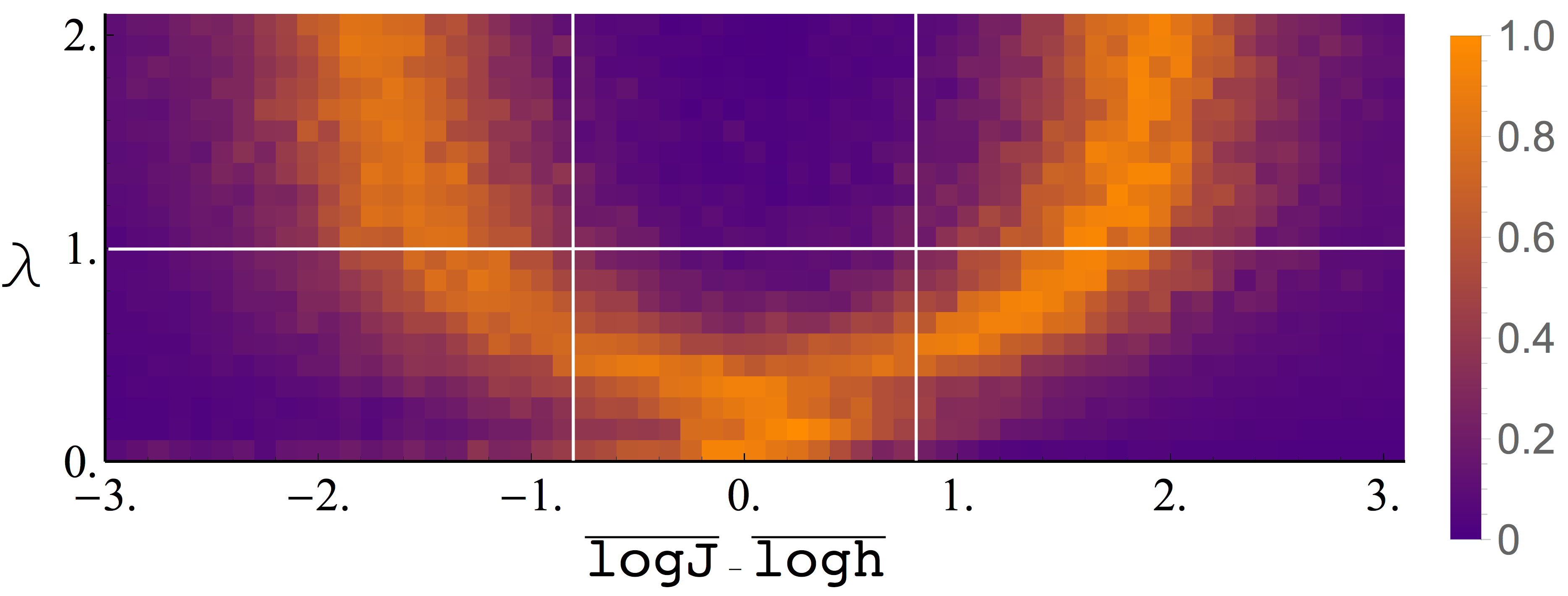}
	\label{fig:conf_std_dev}}
\subfigure[]{
    \includegraphics[width=0.95\linewidth]{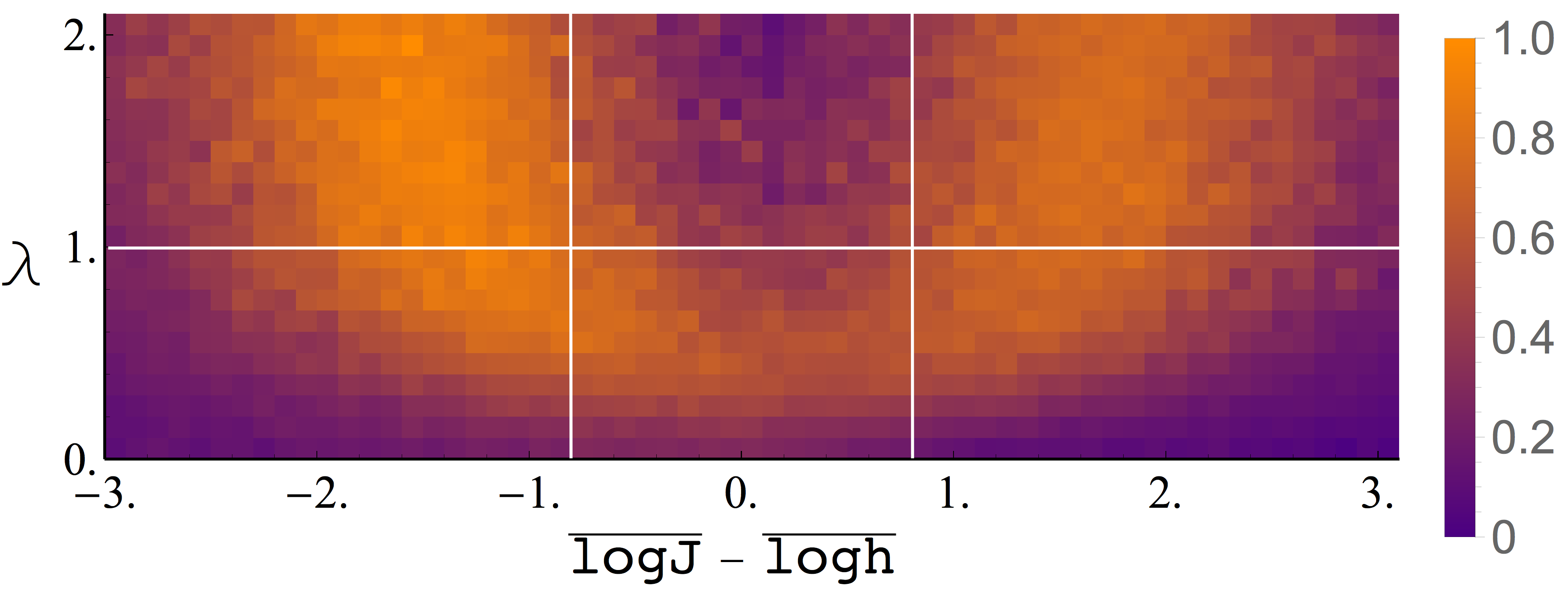}
    \label{fig:EE_std_dev}}
\subfigure[]{
	\includegraphics[width=.34\linewidth]{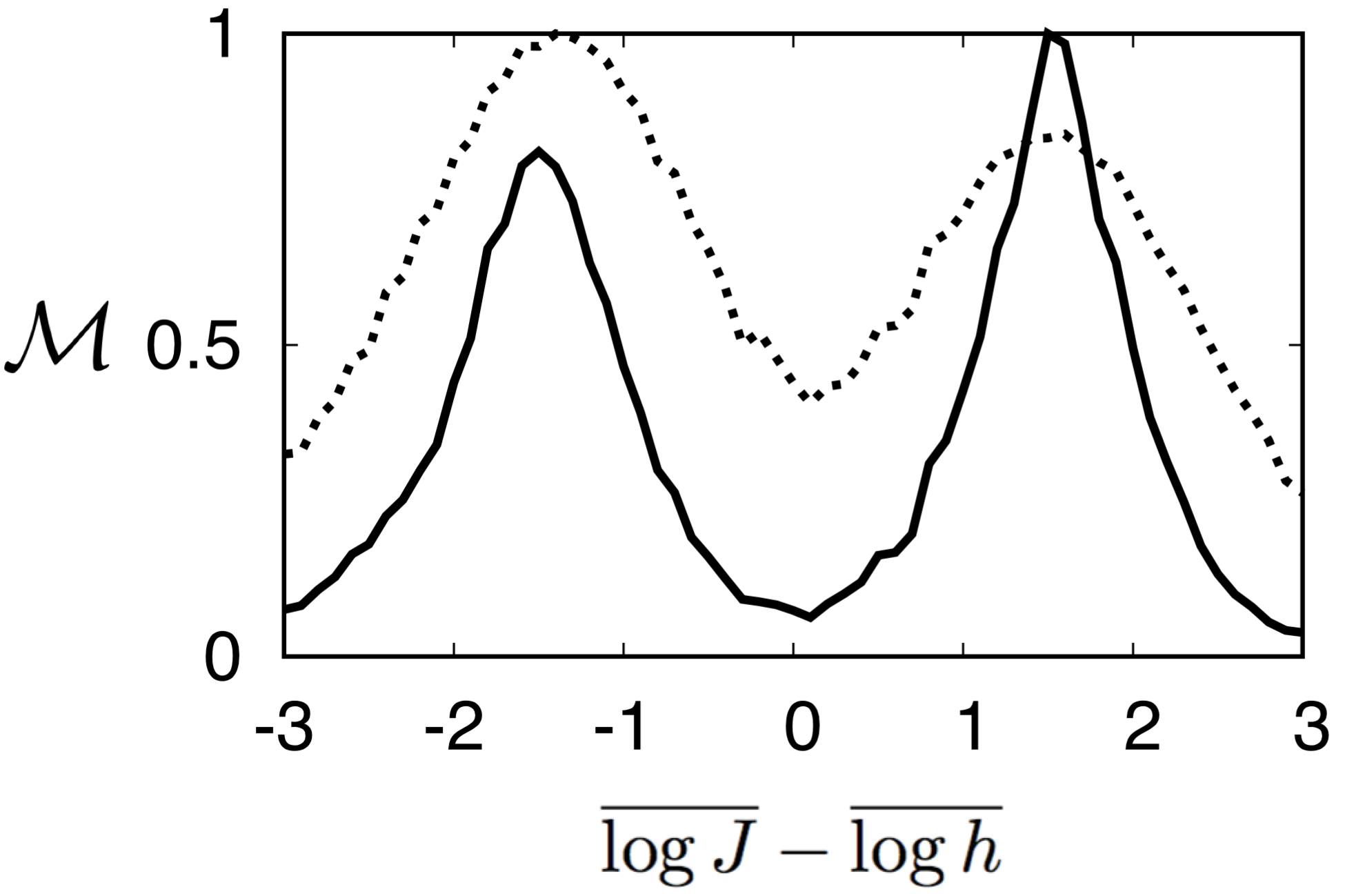}
	\label{fig:conf_std_dev}}
\subfigure[]{
	\includegraphics[width=.28\linewidth]{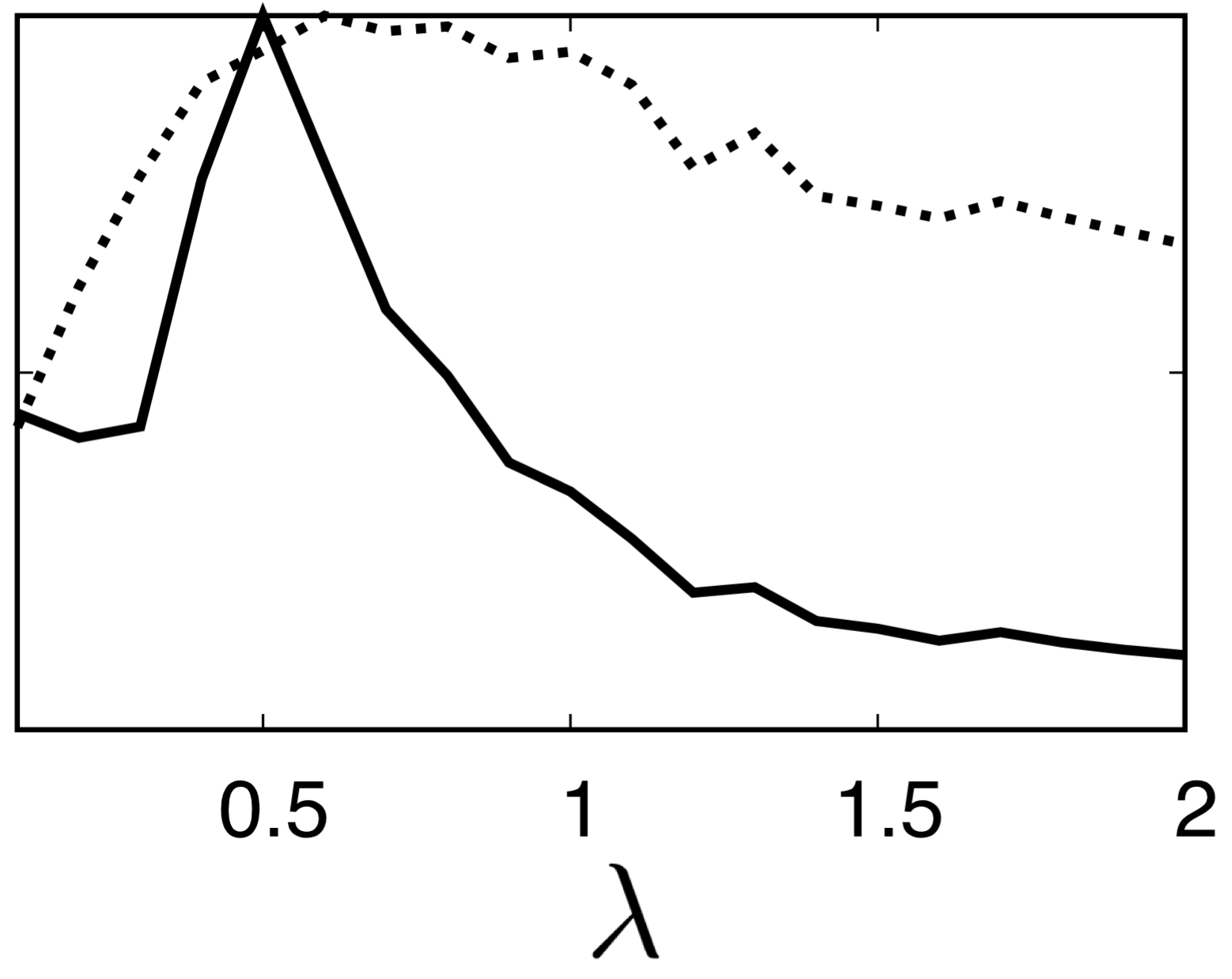}
	\label{fig:conf_std_dev}}
\subfigure[]{
	\includegraphics[width=.28\linewidth]{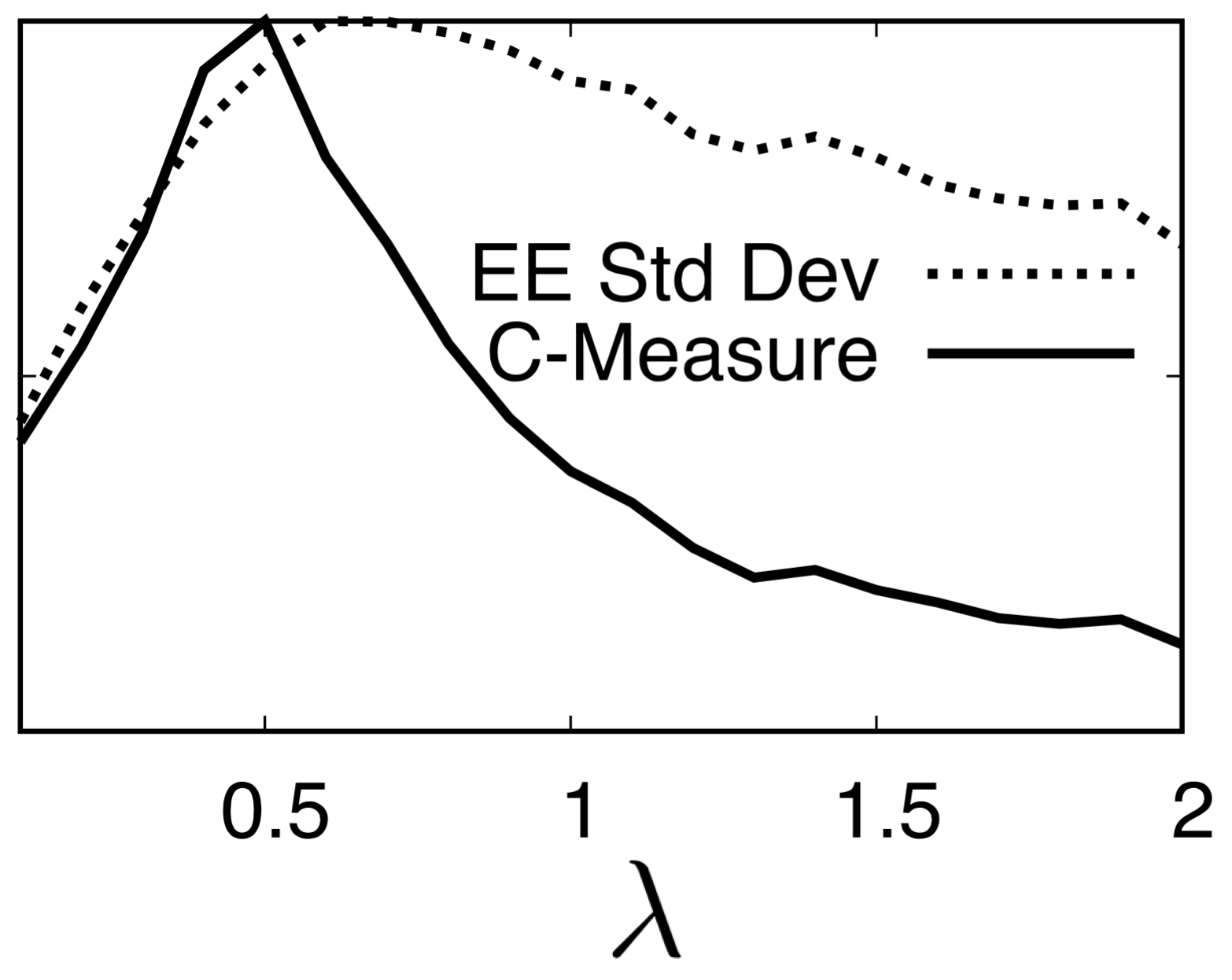}
	\label{fig:conf_std_dev}}
\caption{(a) Our $\mathcal{C}$-measure for extracting phase boundaries (defined in the main text) and (b) the average standard deviation of the entanglement entropy for L=12 using 100 disorder realizations. The data in each has been normalized by the largest value in the parameter space for meaningful comparison.
(c-e) The measures plotted in (a-b) along the cuts marked in white lines:
(c) $\lambda = 1.0$, (d) $\overline{\log J} - \overline{\log h} = +0.8$, and (e) $\overline{\log J} - \overline{\log h} = -0.8$. }
\label{fig:Variance}
\end{figure}

Once $\lambda\neq0$, a numerical study over a large number of disorder realizations looking at all the eigenstates is necessary to obtain the phase diagram that now includes the thermal phase. With finite $\lambda$, the nature and mechanism of various phase transitions largely remain open questions since the existing theoretical understanding is limited to three extreme regimes in the phase space: (1) $J\gg h, \lambda$, (2) $h\gg J,\lambda$, (3)$\lambda\gg J=h$. In the limits (1) and (2), the Anderson localized SG and PM phases of the non-interacting system generalize to MBL versions of themselves\cite{Huse13, PekkerHilbertGlass, Kjall14}. On the other hand, in the strongly interacting limit, the system will be in a thermal phase with its excited states exhibiting volume law entanglement \cite{Page}. Finally, since our interactions were chosen to respect the Ising duality, we expect the phase diagram with non-zero $\lambda$ to still be symmetric about $\overline{\log J} = \overline{\log h}$ (with small corrections for open boundary conditions). Nevertheless, the precise topology of the tri-partite phase boundary and the existence or absence of a direct MBL-MBL phase transition\cite{de2016stability,Moudgalya2017} are hotly debated questions of profound conceptual consequences.
On the other hand, most existing approaches for detecting phase boundaries  relies on the standard deviation of the entanglement entropy (see Fig \ref{fig:Variance}), and these lack sufficient resolution leaving the physics of the critical regime largely inaccessible.   

{\it Neural Network based Approach --} In order to access the information in the entanglement spectra in a holistic manner, we build and employ a feed-forward neural network with a single hidden layer. Our hidden layer contains 200 neurons with sigmoid activation functions. We utilize a cross-entropy cost function with L2 regularization and use a softmax output layer with three neurons, each of which corresponds to one of three possible phases, namely the SG-MBL, the PM-MBL, and the thermal phase (see Fig.~\ref{fig:NN_cartoon}(a)). Since we use a softmax layer, the neuron outputs sum to unity and may be thought of as the probability that a given phase-space point is in a particular phase. The space of possible neuron outputs thus forms a 2-simplex embedded in the three-dimensional space of confidence outputs, with the vertices of the simplex representing points of maximum certainty, see Fig.~\ref{fig:NN_cartoon}(b). 

We then generate the training and testing data for different disorder configurations of the model \eqref{eq:model} on an open chain with 12-sites and open boundary conditions. Specifically, we use exact diagnalization to obtain all the eigenstates and take the middle-quarter of the eigenstates in each Ising symmetry sector to calculate the bipartite entanglement spectra for each eigenstate. 
We generate the training set for three known points of the phase space that correspond to the three target phases: $\overline{\log J} - \overline{\log h} = \pm 0.8$ with $\lambda =0.2$ and $\overline{\log J} - \overline{\log h} = 0.0$ with $\lambda = 1.0$. We use 1000 disorder configurations labeled with each of the three points to train our network to an accuracy of over 90\%. The fact that successful training could be reached already points to the fact that our network could extract and utilize qualitatively distinct information in the entanglement spectra of eigenstates in the three phases of interest.
We found all the results we report below to be insensitive to the exact parameter values used for training (so-called ``hyper-parameters").

\begin{figure}[ht]
\centering
\vspace{10pt}
\subfigure[]{
    \includegraphics[width=0.95\linewidth]{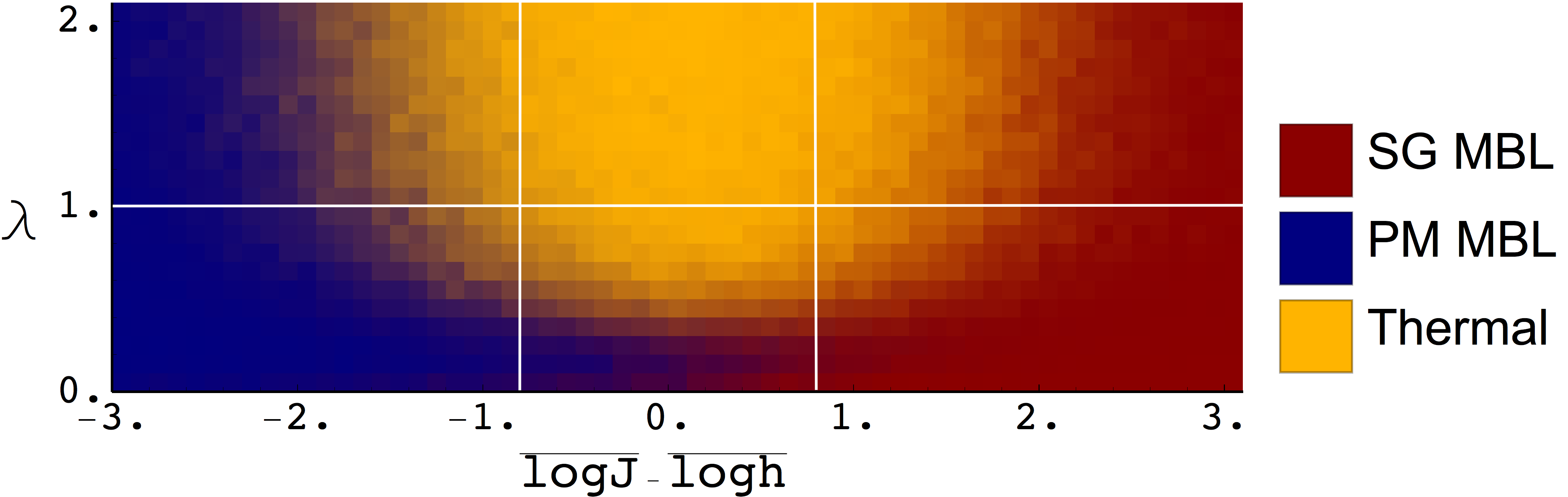}}
\subfigure[]{
    \includegraphics[width=.36\linewidth]{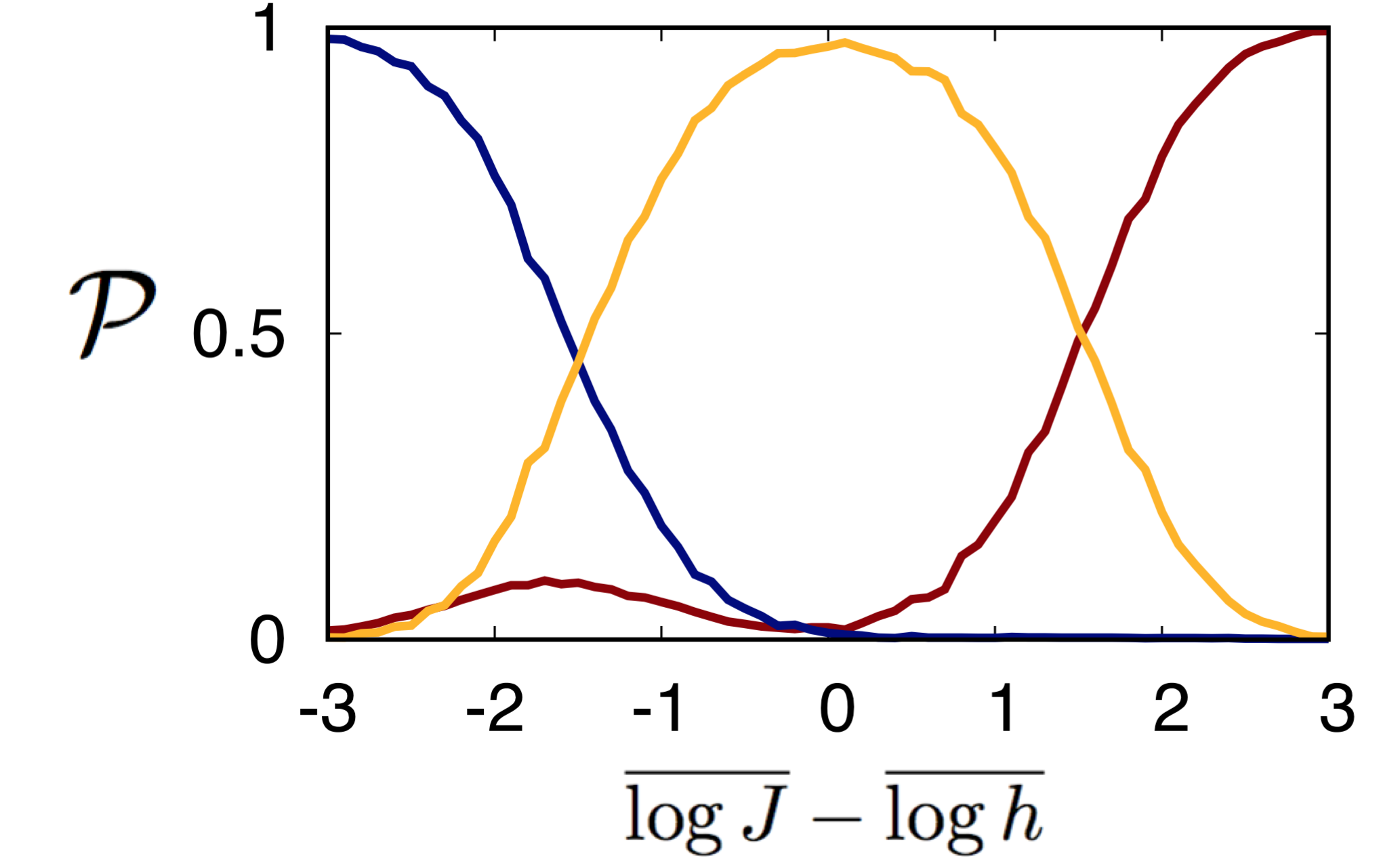}}
\subfigure[]{
    \includegraphics[width=.28\linewidth]{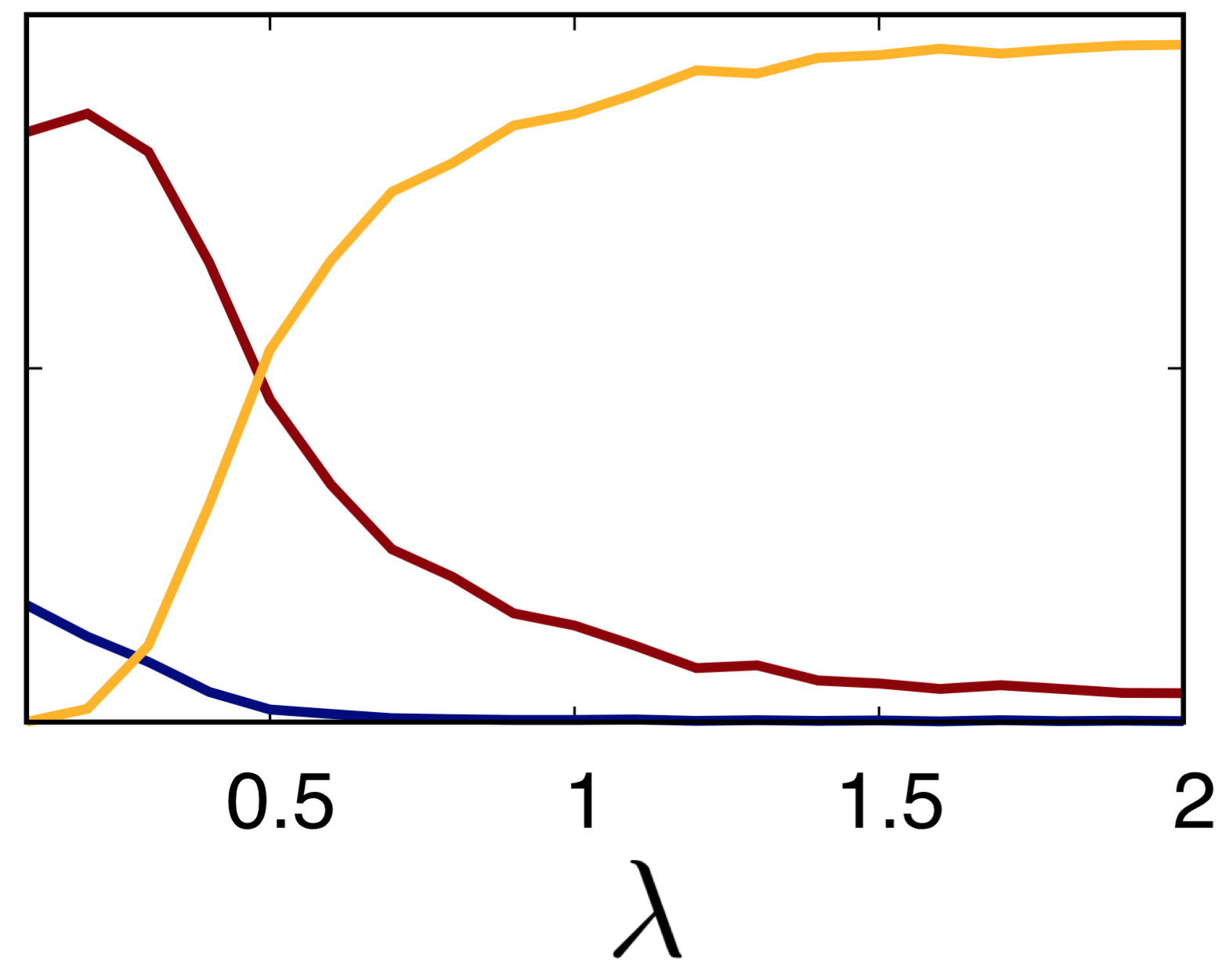}}
\subfigure[]{
    \includegraphics[width=.28\linewidth]{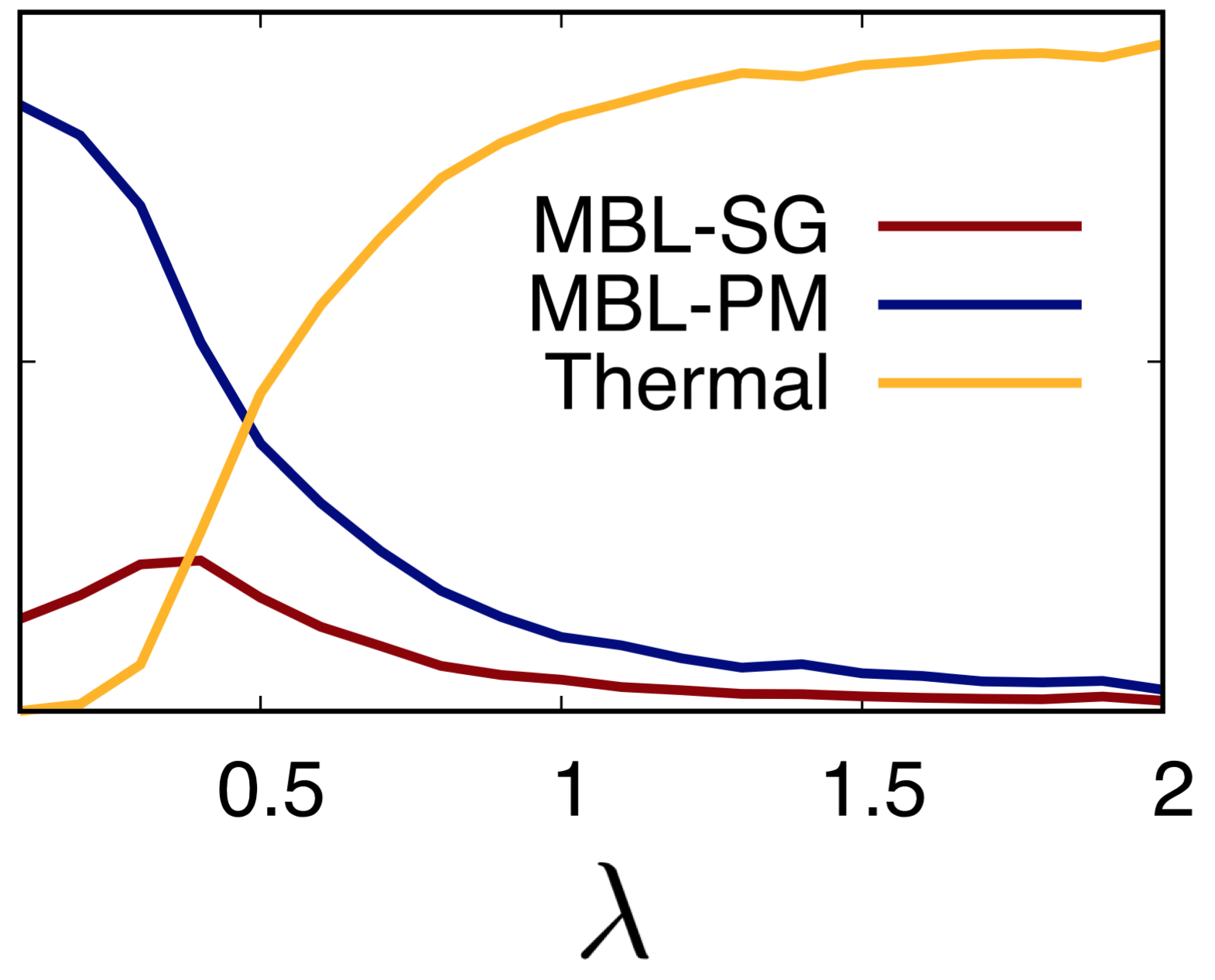}}
\caption{(a) The phase diagram for L=12 using 100 disorder realizations at each point. Here, the average triplet of neural network confidences has been plotted as an RGB parameter. 
In subfigures  
(b-d), the average neural network confidences are plotted along various cuts marked by white lines in (a), obtained using 1000 disorder realizations with L=12. Specifically, (b) $\lambda = 1.0$ (c) $\overline{\log J} - \overline{\log h} = 0.8$ , and (d) $\overline{\log J} - \overline{\log h} = -0.8$  The sampling width is 0.1 for each parameter. }
\label{fig:PhaseDiagram}
\end{figure}

Once the training is complete, we fix the network parameters and feed the entanglement spectra from each point in the phase space of ($\overline{\log J} - \overline{\log h}, \lambda)$ to the network. The network outputs its confidence for the phase space point to belong to one of the three phases (SG-MBL, PM-MBL, Thermal) in the form of a triplet neuron output within the 2-simplex codomain embedded in the three-dimensional space of confidence outputs (see Fig.~\ref{fig:NN_cartoon}(b)). Note that all conventional measures require sampling thousands of disorder configurations. On the other hand, we find that averaging the neuron output over just 100 disorder configurations yields a satisfying phase diagram, paving the way for fast scans of large areas of phase space. The purpose of the averaging is to both look into the statistics, as well as to compare with the conventional measure on equal footing. In Fig.~\ref{fig:PhaseDiagram} (a) we plot the average neural network confidence output in the range of $\overline{\log J} - \overline{\log h} \in [ -3.0, +3.0]$ and $\lambda \in [0.1, 2.0 ]$ by representing each component of the triplet with three colors.  

The phase diagram Fig.~\ref{fig:PhaseDiagram} (a) obtained by the neural network displays several satisfying features that are consistent with theoretical insights. First of all, the phase diagram is  roughly symmetric about the line $\overline{\log J} - \overline{\log h} = 0.0$ and consistent with the Ising duality of the Hamiltonian Eq.\eqref{eq:model}.
Furthermore, the upward curvature of the phase boundary between the MBL phases and the thermal phase is 
consistent with the fact that the non-interacting model is most delocalized near the SG-PM transition\cite{Fisher95} and hence the transition is most susceptible to thermalization upon adding interactions near the $(\overline{\log J} - \overline{\log h} = 0.0, \lambda=0)$ point. 
However, it is evident from the representative line cuts in Figs.~\ref{fig:PhaseDiagram} (b-d) that the variation of the confidence outputs is gradual and broad, masking the precise topology of the phase boundaries.

In order to more precisely study the topology of the phase diagram, we developed a protocol for extracting phase boundaries from multi-neuron outputs. Our approach uses the geometric implication of the fact that neuron outputs sum to unity in a soft-max layer. Specifically, with a soft-max $N$-neuron output, the codomain of the neural network confidence output is a $(N-1)$-simplex embedded in the $N$-dimensional space of outputs.  The points of maximal confusion constitute geometrically notable points on the $(N-1)$-simplex, for $N=3$ these are the mid-points of the edges and the barycenter. 
Explicitly, in our present case the codomain of our neural network is a 2-simplex and the points of maximum confusion that should naturally belong to the phase boundary\cite{Melko2017} are $(1/2, 1/2, 0), (1/2, 0, 1/2), (0, 1/2, 1/2),$ and $(1/3, 1/3, 1/3)$ (see Fig.~\ref{fig:NN_cartoon}). Now for any confidence triplet, one can measure the minimal distance  $d_{\rm min}$ to the set of maximal confusion points. Once we normalize this distance by the maximal possible distance of any point on the simplex to a point of maximal confusion, we obtain 
a continuous measure of confusion capable of extracting boundaries: ${\mathcal{C}}\equiv 1-\bar{d}_{\rm min}$, where $\bar{d}_{\rm min}$ denotes the normalized distance. This measure of confusion ranges between $\mathcal{C}=1$ when the confidence corresponds to one of the maximal confusion points, and $\mathcal{C}=0$ when  
the network outputs a particular phase with 100\% confidence. 

Now at each point in the phase space, we take the average confidence triplet to evaluate the confusion measure $\mathcal{C}$ as shown in Fig.~\ref{fig:Variance}(a). It is notable that the our confusion measure allows us to establish phase boundaries in a manner that is native to the neural network approach. 
Surprisingly, the phase boundary detected by neural network has the topology of a ``wishbone" with a visible phase boundary between two MBL phases (see Fig.~\ref{fig:Variance}(a)) at small $\lambda$. 
This warrants a more exhaustive study of this transition, including finite-size effects in order to probe the existence of a direct SG-MBL to PM-MBL transition.

The ${\mathcal{C}}$-measure based extraction of the phase boundary can be contrasted with a more conventional entanglement entropy based approach\cite{Kjall14}. 
Since the EE changes from area law to volume law upon transition from a MBL phase to a thermal phase, it is expected that the standard deviation of the EE in eigenstates peaks at the phase boundary\cite{Kjall14}. 
Fig.~\ref{fig:Variance}(b) shows the standard deviation taken over all disorder samples and  the middle quarter of the eigenstates from each sample.
As expected, the standard deviation of the EE is peaked at the MBL-thermal boundaries. However, two advantages of the neural-network $\mathcal{C}$ measure easily stand out. First, the EE-based approach cannot distinguish the boundary between the two area-law MBL phases (see the $U$-shaped phase boundary in Fig.~\ref{fig:Variance}(b)) whereas the neural network is successfully differentiating these (see  the ``wishbone" shaped phase boundary in Fig.~\ref{fig:Variance}(a)). For the MBL-SG problem, one can additionally construct an Edwards-Anderson spin-glass order parameter to single out the MBL-SG phase\cite{Kjall14}. However, the ability of the neural network to distinguish between different MBL phases using just the ES and no other ``prior knowledge'' about order-parameters can prove useful for future studies of new MBL phases where order-parameters might be unknown. 
Second, the ${\mathcal{C}}$ measure reveals a markedly sharper phase boundary that enables a better study of its topology (see the line cut comparisons in Fig.~\ref{fig:Variance}(c,d,e)). 

\begin{figure}[ht]
\centering
\includegraphics[width=0.9\linewidth]{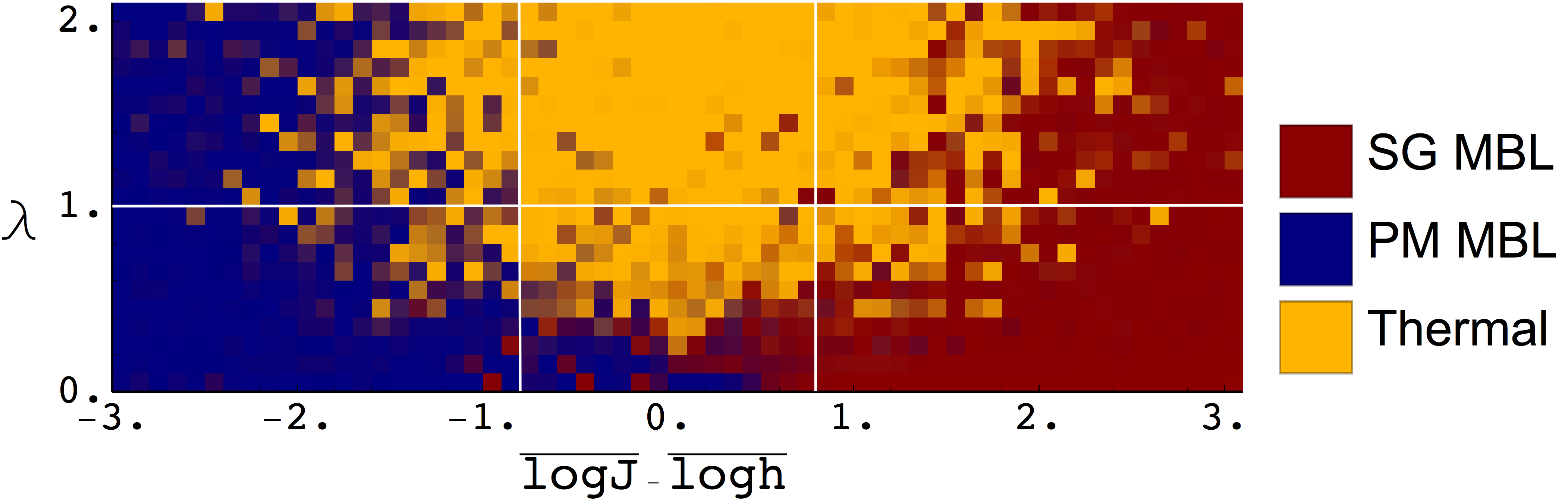}
\caption{The 2D phase diagram where the network has been trained on the full training set but tested on a single disorder realization.}
\label{fig:single}
\end{figure}

Finally, we should remark on the neural networks' ability to see through the noise that is inevitable in studies of disorder effects. Although we have averaged over 100 different disorder configurations to gain statistics in the bulk of this letter,  Fig. \ref{fig:single} shows that the neural network has a remarkable ability to see through the configuration specific noise and capture the coarse features of the phase diagram even for a single disorder realization. The fact that the neural network has gained a regression scheme alternate to the manual modelling of statistical distributions over disorder realizations implies that one can use it as a tool to quickly explore large areas of phase space to map out new non-equilibrium phase diagrams. 

{\it Summary and Outlook --}
Here we exploit the ability of neural networks to distill characteristic features from noisy data in order to extract information from the entanglement spectra associated with out-of-equilibrium phases. To this end, we built a neural network and employed it to process entanglement spectra from a transverse field Ising model with disorder, a poster-child model system that can be in one of three distinct out-of-equilibrium phases. Our neural network, being trained with typical data associated with three limiting points in the phase space, was able to output a phase diagram that is consistent with theoretical expectations. Moreover, using a simplicial geometry construction to quantify network's degree of confusion, we were able to extract the phase boundary with significantly sharper resolution compared to entanglement entropy-based approaches. Any effort to better understand this transition and/or the possibility of an intervening sliver of thermal phase between the two MBL phases will benefit from a method for obtaining a sharper determination of phase boundaries, which our work provides.

The significance of what we have achieved is multi-faceted. First, we have demonstrated that a neural network based approach can give us a sharper look at the multi-partite phase boundary by using the geometric measure of confusion $\mathcal{C}$ that we introduced. This is the first example, to the best of our knowledge, that a neural network based approach outperformed conventional approach in terms of sharper phase boundaries. Our work paves the way for future studies on the much debated topic of the nature of MBL phase transitions. Second, by having multiple neuron outputs, we were able to a obtain tripartite phase diagram involving two distinct MBL phases with a \emph{single} measurement. This is valuable even for MBL phases where there are known order-parameters~\cite{Kjall14} as in the model we considered. 
However, this multi-neuron output approach will be even more valuable when dealing with new out-of-equilibrium phases without \emph{a priori} knowledge of suitable order parameters. 

{\it Acknowledgements:} E-AK and JV thank Yi Zhang for discussions. 
 E-AK acknowledges the Simons Fellow
in Theoretical Physics Award \#392182 and DOE support
under Award de-sc0010313. E-AK is grateful to
the hospitality of the Kavli Institute of Theoretical Physics
supported by NSF under Grant No. NSF PHY-1125915,
where this work was initiated. JV acknowledges NSF support
 under Award NSF DMR-1308089.
VK thanks S. Moudgalya and D. Huse for an ongoing collaboration on the model studied in this paper. VK is supported by the Harvard Society of Fellows and the William F. Milton Fund. 

\bibliography{global}

\end{document}